\documentstyle[12pt,epsf]{article}

\begin{document}

\begin{titlepage}

\null
\begin{flushright}
CERN-TH.7197/94\\
ILL-(TH)-94-07
\end{flushright}
\vspace{20mm}

\begin{center}
\bf\Large
On the absence of an exponential bound in\\
four dimensional simplicial gravity
\end{center}

\vspace{5mm}

\begin{center}
{\bf S. Catterall
\footnote{Permanent address: Physics Department,
Syracuse University, Syracuse, NY 13244}}\\
TH-Division, CERN CH-1211,\\
Geneva 23, Switzerland.\\ 
{\bf J. Kogut}\\
Loomis Laboratory, University of Illinois at Urbana,\\
1110 W. Green St, Urbana, IL 61801.\\
{\bf R. Renken}\\
Department of Physics, University of Central Florida,\\
Orlando, FL 32816.
\end{center}
      
\vspace{10mm}
 
\begin{abstract}
We have studied a model which has been
proposed as a regularisation for four dimensional quantum gravity.
The partition function is constructed by performing a
weighted sum over all triangulations of the four sphere. 
Using numerical simulation we find that the number of such
triangulations containing $V$ simplices grows faster than
exponentially with $V$. This property ensures that the model
has no thermodynamic limit. 
\end{abstract}

\vfill
\begin{flushleft}
CERN-TH.7197/94\\
March 1994
\end{flushleft}

\end{titlepage}

\section*{Introduction}

In the last few years there has been considerable interest generated
in a model for quantum gravity in which the functional integral
over metrics (ill-defined in the continuum) is replaced by a
discrete sum over random triangulations. The initial proposal
\cite{mig,amb,gross} arose as a natural generalisation of random
surface theories in two dimensions. The results of these numerical
studies were encouraging and were confirmed by other groups 
\cite{varsted,brug}. The most exciting possibility was the observation
of a possible phase transition for a critical value of the bare
Newton constant. The hope was that a nonperturbative quantum
theory for gravity could be recovered in the vicinity of this
new fixed point. These observations were rendered more
quantitative by the recent work \cite{us} in which a serious
finite size scaling study was performed.    

The model is defined from the
partition function.
\begin{equation}
Z=\sum_{T\left(S^4\right)} e^{-\kappa_4 N_4 + \kappa_0 N_0}
\label{eqn1}
\end{equation}
The sum is restricted to run over all simplicial manifolds 
(triangulations) with
the topology of $S^4$. 
The first term in the action $N_4$ is just the number of four simplices
in the triangulation $T$ and this allows us to identify the corresponding
coupling $\kappa_4$ as a bare cosmological constant. The second term 
depends only on the number of vertices in the
triangulation $N_0$ and plays the 
role of the integrated Ricci scalar -- the coupling $\kappa_0$ is then
essentially the
inverse bare Newton constant. 

This correspondence is clear classically from
the usual Regge expression for the curvature associated to
any triangle $r_{ijk}$ 
with the extra constraint that the four simplices are all
considered equilateral
\begin{equation}
r_{ijk}=2\pi - \cos^{-1}{\left(1\over 4\right)} n_4^{ijk}
\end{equation}
Notice that if the volume is bounded, the number of four simplices
shared by a given triangle $n_4^{ijk}$ is necessarily also bounded.
This automatically ensures that
the model is well defined at finite volume -- it is a dynamical
question as to whether the problems associated to the 
unboundedness of the continuum action return on taking the
large volume limit.

As we have remarked, the analogous model in two dimensions has been 
studied extensively, 
see, for example, the review \cite{david}. It seems clear
that at least for central charges less than unity, the sum over triangulated
graphs correctly mimics the continuum functional integrals over
the metric including the
conformal anomaly. In four dimensions it is not at all clear that a simple
generalisation, such as the one described above, is sufficient to explore
the space of metrics.
However, it constitutes a
simple ansatz which may be studied using numerical simulation.

We may rewrite eqn.\ref{eqn1} in the form
\begin{equation}
Z=\sum_{N_4}e^{-\kappa_4 N_4} \Omega\left(N_4,\kappa_0\right)
\end{equation}
The partial sum $\Omega\left(N_4,\kappa_0\right)$ counts the number
of triangulations (weighted by the Ricci term) with volume $N_4$. The
results we discuss here are concerned with the volume dependence 
of this entropy function $\Omega\left(N_4,\kappa_0\right)$. 
It is helpful at this point to recall the behaviour of
the equivalent two dimensional model.

Two dimensional gravity regulated using dynamical triangulations has 
a partition analogous to eqn. \ref{eqn1}. The number of simplices
is now just $N_2$ with corresponding cosmological constant $\kappa_2$.
The coupling $\kappa_0$ plays no role in two dimensions as the 
number of vertices $N_0$ is strictly proportional to the 
number of two simplices $N_0={1\over 2}N_2+\chi$ 
if the Euler character $\chi$ is kept constant (for
example $S^2$). 
If we sum over all two dimensional triangulations
with fixed volume we arrive at a quantity $\omega\left(N_2\right)$
analogous to
$\Omega\left(N_4,0\right)$ for the four dimensional theory.

There are rigorous proofs \cite{tutte,bessis}
that this quantity is exponentially
bounded.
\begin{equation}
\omega\left(N_2\right)\sim e^{\kappa_2^c N_2}
\end{equation}
This property is {\it crucial} for the very existence of the
partition function. It implies that for a sufficiently large bare 
cosmological constant
$\kappa_2 > \kappa_2^c$
the partition function will be finite. The thermodynamic limit
$N_2 \to\infty$ is then obtained by tuning $\kappa_2$ towards this critical
value $\kappa_2^c$. The mean volume $\left\langle N_2\right\rangle$ then
behaves as $\left\langle N_2\right\rangle\sim \left(\kappa_2-\kappa_2^c
\right)^{-1}$.
If the number of triangulations were to increase faster than
exponentially, it would be impossible to tune the bare cosmological
constant to approach the large volume limit in a regular fashion -- the
partition function would be dominated by infinite volume triangulations
independent of the bare lattice parameters. Constructing a 
continuum limit would then be impossible.

Thus, it is absolutely essential for the very existence of these higher
dimensional models that there be such a bound.
Unfortunately, there are no analytic proofs available for dimension
greater than two. If the topology is not fixed it
can be shown that the number of triangulations increases
factorially with volume even in
two dimensions \cite{amb3}. The situation is made worse by the
lack of any topological classification of three and four 
dimensional manifolds. 

Faced with this we have used numerical simulation to 
estimate the volume of the
triangulation space. Whilst the previous studies \cite{mig,amb,varsted,brug}
have claimed evidence
for an exponential bound we believe the issue is of such paramount
importance that a very detailed study is required. Indeed, the results we 
shall present
favour a very different scenario.

\section*{Method}

For an entropy function that behaves exponentially with volume 
we have argued that it is 
possible to
choose the coupling $\kappa_4$ to fix the mean volume $\left\langle
N_4 \right\rangle$. In practice this
is a difficult fine tuning problem. Even under the
assumption of an exponential bound, the entropy
$\Omega\left(N_4,\kappa_0\right)$ is of the form

\begin{equation}
\Omega\left(N_4,\kappa_0\right)\sim N_4^
{a\left(\kappa_0\right)}
e^{\kappa_4^c\left(\kappa_0\right) N_4} 
\end{equation}

We have included the leading power law correction parametrised
by $a\left(\kappa_0\right)$. In practice the power $a$ is negative, so that the
partition function is dominated by small or large volumes depending on
the sign of
$\Delta\kappa_4=\kappa_4-\kappa_4^c\left(\kappa_0\right)$. 

This problem has been tackled in a variety of
ways. We have followed the approach of Migdal et al. \cite{mig} and
added to the action a small correction term of the form
\begin{equation}
\Delta S=\gamma\left(N_4-V\right)^2
\end{equation}
Replacing the sums by integrals and forgetting for the moment any
power law corrections 
it is now simple to obtain a relation between the
mean volume $\left\langle N_4\right\rangle$ and the parameters
in the action.
\begin{equation}
\left\langle N_4\right\rangle={1\over 2\gamma}\left(\Delta\kappa_4+
2\gamma V\right)
\label{iter}
\end{equation}
Thus tuning $\kappa_4$ to yield an average volume $V$ yields a
measurement of the coupling $\kappa_4^c\left(\kappa_0\right)$. 
The auxiliary coupling $\gamma$ 
merely controls the magnitude of volume fluctuations. We have
set $\gamma=0.005$.
The presence of power law (and other
subleading) corrections gives $\kappa_4^c\left(\kappa_0\right)$ 
a volume dependence
$\kappa_4^c\left(\kappa_0\right)=\kappa_4^c\left(V,\kappa_0\right)$. 
The relation eqn. \ref{iter}
may be rewritten
\begin{equation}
\kappa_4^c\left(N_4,\kappa_0\right)=\kappa_4+2\gamma\left(\left\langle N_4
\right\rangle - V\right)
\end{equation}

In practice we iterate the above relation during
the thermalisation stage of our simulation and apply it once more at the
end of our run to
compute our final estimate for $\kappa_4^c$. 
 
In this picture the presence of an exponential bound would be signalled by
this critical cosmological constant
$\kappa_4^c\left(V,\kappa_0\right)$ having a finite limit for large volumes $V$.
In contrast 
$\kappa_4^c\left(V,\kappa_0\right)$ would increase
logarithmically in a model 
for which $\Omega\left(N_4,\kappa_0\right)$ grew
factorially with volume (this just follows from the asymptotic result
$\left(x!\right)^\delta\sim e^{\delta x\ln x}$). 

Notice that it is
sufficient to prove an exponential bound for a single value of
$\kappa_0$ -- the following inequality guarantees that there will then be
a bound for any other $\kappa_0 > 0$.
\begin{equation}
\Omega\left(N_4,0\right)\le\Omega\left(N_4,\kappa_0\right)\le
\exp\left(\alpha\kappa_0 N_4
\right)\Omega\left(N_4,0\right)
\label{ineq}
\end{equation}

We have used a Monte Carlo algorithm to sample the triangulation space
of the model -- the details are given in
\cite{us2}. 
Our code is written in such a way as to make the
dependence on dimension $d$ trivial -- it enters only as an input parameter
to the program. 

We have simulated systems from size $V=500$ to $V=32000$.
Typical runs utilised on the
order of $4\times 10^5$ MC sweeps with one sweep corresponding to
$V$ trial updates.
In addition we performed a series of runs for both the two dimensional
and three dimensional models. The results of these simulations could
then be contrasted with the equivalent four dimensional data and
served as an important test of our code.

\section*{Results} 

\begin{figure}
\begin{center}
\leavevmode
\epsfxsize=400pt
\epsfbox{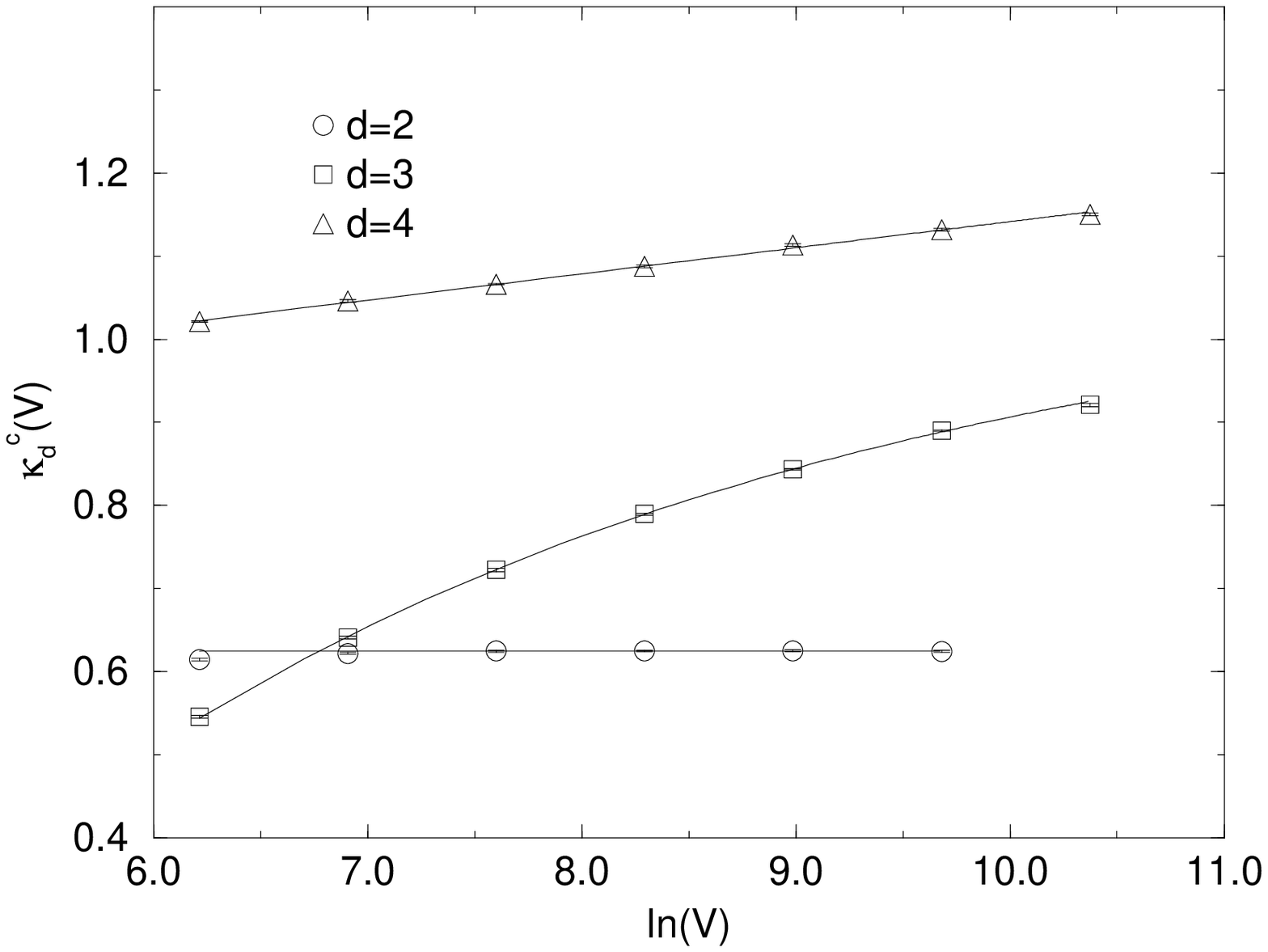}
\caption{Critical cosmological constant $\kappa_0=0$}
\label{fig1}
\end{center}
\end{figure}

Fig.\ \ref{fig1} is a plot of the critical cosmological coupling
$\kappa_d^c\left(V,\kappa_0\right)$ against the logarithm of the volume
for the two, three and four dimensional models at $\kappa_0=0$
(To improve clarity we plot $\kappa_2^c-0.5$ and $\kappa_3^c-1.0$).
Clearly, the presence of an exponential bound emerges very clearly
in the two dimensional case -- $\kappa_2^c\left(V,0\right)$ is
statistically consistent with a constant $\kappa_2^c\left(\infty\right)=
1.1249(6)$ for volumes $V\ge 2000$.

For three dimensions the situation is rather different. The finite
volume dependence of $\kappa_3^c\left(V,0\right)$ is large over the
full range of volumes analysed. However, as the plot reveals there is
no strong evidence of a logarithmic component -- indeed the best fit we
could make to the data corresponds to a {\it convergent} power
law (the solid line in the figure) $\kappa_3^c=a+bV^c$.
The fit yields $a=2.01(1)$, $b=-3.2(1)$ and $c=-0.28(1)$ with a $\chi^2$
per degree of freedom $2.0$. Thus, our data in three dimensions favours
a bound. Indeed these numbers are consistent with the ones quoted
in a previous study by Ambj\o rn and Varsted \cite{ambvar} who give
$a=2.06$, $b=-3.9$ and $c=-0.32$. Their fit derives from lattice
sizes of $V=14000$ and smaller with lower statistics but it is
reassuring to see that we are in pretty good agreement. We are
currently extending our dimension three runs to larger lattices to
strengthen our confidence in the three dimensional bound.

The situation in four dimensions is radically different. 
Clearly, the data
support the hypothesis that there is a logarithmic component to
the critical volume coupling $\kappa_4^c\left(V,0\right)$.
A fit of all the $d=4$ data to a simple logarithm $\kappa_4^c=a+b\ln V$
results in a value for $b=0.0315(3)$ with a $\chi^2$ per degree of 
freedom $2.7$ (solid line shown).
Converging power fits simply fail to describe the data. 

\begin{figure}
\begin{center}
\leavevmode
\epsfxsize=400pt
\epsfbox{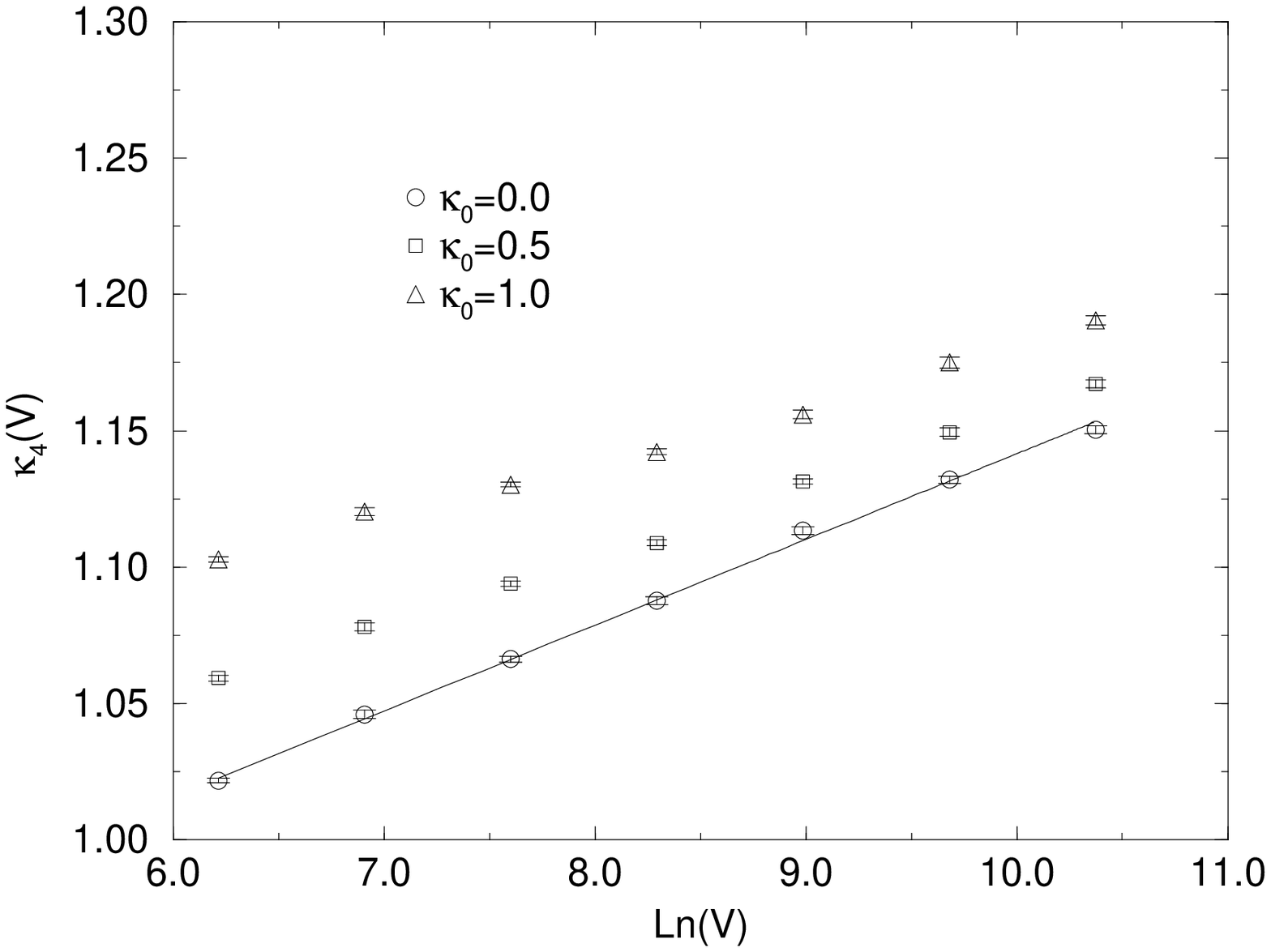}
\caption{$\kappa_4^c\left(V\right)$ $d=4$ }
\label{fig2}
\end{center}
\end{figure}

To test this hypothesis further we looked at the situation for
non zero $\kappa_0$. Fig.\ \ref{fig2} shows a plot of $\kappa_4^c\left(V,
\kappa_0\right)$ for $\kappa_0=0.0$, $\kappa_0=0.5$ and $\kappa_0=1.0$.
The inequality eqn.\ref{ineq} implies that the coefficient of this logarithm
should be universal (independent of $\kappa_0$). 
The leading effect of a
non zero value for $\kappa_0$ is simply a renormalisation of any
exponential terms in $\Omega\left(N_4,0\right)$. This is confirmed by
the data in fig.\ \ref{fig2}. Although the curves start out with
different gradients their large volume behaviour appears to be
the same.

However the plot makes it
also clear that the onset of this asymptotic regime is dependent on
$\kappa_0$ -- as $\kappa_0$ increases the curves start off
increasingly flat and the logarithm only manifests itself for
large volumes. 

We found that very long runs were required to thermalise the four
dimensional lattices. 
The initial configurations were created by employing only
the node insertion move which effectively generates lattices corresponding
to large values of $\kappa_0$. For the largest volumes we employed, $V=32000$,
we found that subsequent relaxation times were of the order of $10^5$ sweeps.
This difficulty of reaching true equilibrium was the main
factor in determining the largest volumes we could reach. It is perhaps
a practical demonstration of the results reported in \cite{radi} in
which the algorithmic unrecognisability of four manifolds is shown
to lead to a lack of a reasonable bound on the number of local moves
needed to pass from one configuration to another.

Thus our four dimensional data would indicate that the
entropy $\Omega\left(V,0\right)$ has a leading behaviour
\begin{equation}
\Omega\left(V,0\right)\sim \left(V!\right)^\delta
\end{equation}

If we fit the $\kappa_0=0.0,\, 0.5,\, 1.0$ data for the three largest volumes
by straight lines we find consistent estimates for the exponent $\delta$.
These are $\delta=0.027(1),\, 0.026(1),\, 0.025(2)$ respectively.
We would then assign our best estimate for $\delta$ as $\delta=0.026(5)$.

\section*{Outlook}

In summary, we have presented results which are consistent with a 
leading factorial
behaviour for the entropy of triangulations of the four sphere
$\Omega\left(V,0\right)$. Specifically,
the number of triangulations of $S^4$ grows like
\begin{equation}
\Omega\left(V,0\right)\sim \exp\left(aV\right)\left(V!\right)^\delta
\end{equation}
Furthermore, we estimate the exponent $\delta=0.026(5)$.
This rapid growth renders it impossible to take the thermodynamic
(large volume) limit -- the partition function for any $\kappa_0$
is dominated by large volumes. This in turn implies there is
no continuum limit for the model.

We have argued that the presence of large finite volume effects
can obscure this behaviour for large values of the 
inverse bare Newton constant ($\kappa_0$)
on lattices that are computationally accessible. It is tempting to
speculate that the rather rapid shift of the pseudo critical node coupling
reported in \cite{us} is further evidence for the lack of a
well-defined continuum limit. The data presented in \cite{us} is not
inconsistent with a scenario in which this pseudo-critical coupling
diverges as the mean volume approaches infinity, leaving the system
in an extremely crumpled, degenerate phase.

It is important to notice also that the term added to help fine tune 
the cosmological
constant $\kappa_4$ is
now playing a crucial role in defining the partition function. There is
now no reason to believe that different methods of doing this are
equivalent. 

Clearly an extension of this work (with perhaps a more refined method
for computing $\kappa_4^c\left(V,\kappa_0\right)$) to larger 
volumes and node couplings would
help to confirm these conclusions.

\section*{Acknowledgements}
 
These calculations were supported in part by NSF grant PHY92-00148.
SMC acknowledges useful discussions with M. Bauer, T. Morris and
A. Shapere.  Some calculations were performed on the Florida State University
Cray Y-MP.

\vfill
\newpage

\vfill
\newpage

\vfill
\newpage

\begin{thebibliography}{99}
\bibitem{mig} M. Agishtein and A. Migdal, Nucl. Phys. B385 (1992) 395.
\bibitem{amb} J. Ambj\o rn and J. Jurkiewicz, Phys. Lett. B278 (1992) 42. 
\bibitem{gross} N. Godfrey and M. Gross, Phys. Rev. D43 (1991) R1749.
\bibitem{varsted} S. Varsted, UCSD/PTH 92/03. 
\bibitem{brug}  B. Brugmann and E. Marinari, Phys. Rev. Lett 70 (1993) 1908. 
\bibitem{us} S. Catterall, J. Kogut and R. Renken, CERN-TH.7149/94, 
ILL-(TH)-94-26. submitted to Phys. Lett. B.
\bibitem{david} F. David, `Simplicial Quantum Gravity and Random Surfaces',
Saclay preprint T93/028.
\bibitem{tutte} W.T. Tutte, Can. J. Math. 14 (1962) 21.
\bibitem{bessis} D. Bessis, C. Itzykson and J. Zuber, Adv. Appl. Math. 1 (1980)
109.
\bibitem{amb3} J. Ambj\o rn, B. Durhuus and T. Jonsson, Mod. Phys. Lett. A 12
(1991) 1133.
\bibitem{us2} S. Catterall, J. Kogut and R. Renken, in preparation.
\bibitem{ambvar} J. Ambj\o rn and S. Varsted, 'Entropy estimate
in three dimensional simplicial quantum gravity', NBI-HE-91-17.
\bibitem{radi} A. Nabutovsky and R. Ben-Av, Commun. Math. Phys. 157 (1993) 93.
\end{thebibliography}
\end{document}